\documentclass [12pt,eqsecnum,amsfonts,aps]{revtex4}

\newcommand{\beq}{\begin{equation}}
\newcommand{\eeq}{\end{equation}}
\newcommand{\beqs}{\begin{eqnarray}}
\newcommand{\eeqs}{\end{eqnarray}}
\newcommand{\lsim}{\mathrel{\raisebox{-
.6ex}{$\stackrel{\textstyle<}{\sim}$}}}
\newcommand{\gsim}{\mathrel{\raisebox{-
.6ex}{$\stackrel{\textstyle>}{\sim}$}}}

\begin{document}

\baselineskip 6.0mm

\title{Condensates in Quantum Chromodynamics and the Cosmological Constant}

\author{Stanley J. Brodsky$^{a,b}$}
%\thanks{sjbth@slac.stanford.edu}

\author{Robert Shrock$^b$}
%\thanks{robert.shrock@sunysb.edu}

\affiliation{(a)
Stanford Linear Accelerator Center, Stanford University, Stanford, CA  94309}

\affiliation{(b)
C.N. Yang Institute for Theoretical Physics, Stony Brook University,
Stony Brook, NY 11794}

\begin{abstract}

Casher and Susskind have noted that in the light-front description,
spontaneous chiral symmetry breaking in quantum chromodynamics (QCD) is a
property of hadronic wavefunctions and not of the vacuum.  Here we show from
several physical perspectives that, because of color confinement, quark and
gluon QCD condensates are associated with the internal dynamics of hadrons.  We
discuss condensates using condensed matter analogues, the AdS/CFT
correspondence, and the Bethe-Salpeter/Dyson-Schwinger approach for bound
states.  Our analysis is in agreement with the Casher and Susskind model and
the explicit demonstration of ``in-hadron'' condensates by Roberts et al.,
using the Bethe-Salpeter/Dyson-Schwinger formalism for QCD bound states.  These
results imply that QCD condensates give {\it zero} contribution to the
cosmological constant, since all of the gravitational effects of the in-hadron
condensates are already included in the normal contribution from hadron masses.

\end{abstract}


\maketitle

\pagestyle{plain}
\pagenumbering{arabic}

\section{Introduction}

Hadronic condensates play an important role in quantum chromodynamics.  Two
important examples are $\langle \bar q q \rangle \ \equiv \ \langle
\sum_{a=1}^{N_c} \bar q_a q^a \rangle$ and $\langle G_{\mu\nu}G^{\mu\nu}
\rangle \ \equiv \ \langle \sum_{a=1}^{N_c^2-1} G^a_{\mu \nu} G^{a \, \mu\nu}
\rangle$, where $q$ is a light quark (i.e., a quark with current-quark mass
small compared with the QCD scale $\Lambda_{QCD}$), $G^a_{\mu\nu} =
\partial_\mu A^a_\nu - \partial_\nu A^a_\mu + g_s c_{abc} A_\mu^b A_\nu^c$,
$a,b,c$ are color indices, and $N_c=3$.  (For most of the paper we focus on QCD
at zero temperature and chemical potential, $T==\mu=0$.)  For QCD with $N_f$
light quarks, the $\langle \bar q q \rangle = \langle \bar q_{_L} q_{_R} + \bar
q_{_R} q_{_L} \rangle$ condensate spontaneously breaks the global chiral
symmetry ${\rm SU}(N_f)_L \times {\rm SU}(N_f)_R$ down to the diagonal,
vectorial subgroup ${\rm SU}(N_F)_{diag}$, where $N_f=2$ (or $N_f=3$ since $s$
is a moderately light quark).  Thus in the usual description, one identifies
$\langle \bar q q \rangle \sim \Lambda_{QCD}^3$ and $\langle
G_{\mu\nu}G^{\mu\nu} \rangle \sim \Lambda_{QCD}^4$, where $\Lambda_{QCD} \simeq
300$ MeV.  These condensates are conventionally considered to be properties of
the QCD vacuum and hence to be constant throughout spacetime.  A consequence of
the existence of such vacuum condensates is contributions to the cosmological
constant from these condensates that are $10^{45}$ times larger than the
observed value.  If this disagreement were really true, it would be an
extraordinary conflict between experiment and the Standard Model.

A very different perspective on QCD condensates was first presented in a
seminal paper by Casher and Susskind~\cite{cs} published in 1974. These authors
argued that ``spontaneous symmetry breaking must be attributed to the
properties of the hadron's wavefunction and not to the vacuum''~\cite{cs}. The
Casher-Susskind argument is based on the Weinberg's
infinite-momentum-frame~\cite{Weinberg:1966jm} Hamiltonian formalism of QCD,
which is equivalent to light-front (LF) quantization and Dirac's front
form~\cite{Dirac:1949cp} rather than the usual instant form.  Casher and
Susskind also presented a specific model in which spontaneous chiral symmetry
breaking occurs within the confines of the hadron wavefunction due to a phase
change supported by the infinite number of quark and quark pairs in the
light-front wavefunction.  In fact, the Regge behavior of hadronic
structure functions requires that LF Fock states of hadron have Fock states
with an infinite number of quark and gluon partons~\cite{Mueller:1993rr,
Antonuccio:1997tw,Motyka:2009gi}.  Thus, in contrast to formal discussions in statistical
mechanics, infinite volume is not required for a phase transition in
relativistic quantum field theories.

Spontaneous chiral symmetry breaking in QCD is often analyzed by means of an
approximate solution of the Schwinger-Dyson equation for a massless quark
propagator; if the running coupling $\alpha_s=g_s^2/(4\pi)$ exceeds a value of
order 1, this yields a nonzero dynamical (constituent) quark mass
$\Sigma$~\cite{sd}. Since in the path integral, $\Sigma$ is formally a source
for the operator $\bar qq$, one associates $\Sigma \ne 0$ with a nonzero quark
condensate.  However, the Dyson-Schwinger equation, by itself, does not
incorporate confinement and the resultant property that quarks and gluons have
maximum wavelengths~\cite{lmax}; further, it does not actually determine where
this condensate has spatial support or imply that it is a spacetime constant.

In contrast, let us consider a meson consisting of a light quark $q$ bound to a
heavy antiquark, such as a $B$ meson.  One can analyze the propagation of the
light quark $q$ in the background field of the heavy $\bar b$ quark.  Solving
the Dyson-Schwinger equation for the light quark, one obtains a nonzero
dynamical mass and thus a nonzero value of the condensate $\langle \bar q q
\rangle$.  But this is not a true vacuum expectation value; instead, it is the
matrix element of the operator $\bar q q$ in the background field of the $\bar
b$ quark; i.e., one obtains an {\it ``in-hadron"} condensate.

The concept of ``in-hadron condensates" was in fact established in a series of
pioneering papers by Roberts et
al.~\cite{Maris:1997hd,Maris:1997tm,Langfeld:2003ye} using the
Bethe-Salpeter/Dyson-Schwinger analysis for bound states in QCD in conjunction
with the Banks-Casher relation $-\langle \bar q q \rangle = \pi \rho(0)$, where
$\rho(\lambda)$ denotes the density of eigenvalues $\pm i\lambda$ of the
(antihermitian) Euclidean Dirac operator \cite{bcrel}.  These authors
reproduced the usual features of spontaneous chiral symmetry breaking using
hadronic matrix elements of the Bethe-Salpeter eigensolution.
For example, as shown by Maris, Roberts and Tandy,\cite{Maris:1997hd}
the Gell Mann-Oakes-Renner relation~\cite{gmor}  for a pseudoscalar hadron in the Bethe-Salpeter analysis is
$f_H m^2_H = - \rho^H_\zeta {\cal M}_H,$
where ${\cal M}_H$    is the  sum of current quark masses and $ f_H$ is  the meson decay constant:
\beq
f_H P^\mu = Z_2 \int^\Lambda \frac{d^4q}{(2\pi)^4}
 ~{1\over 2} \big[ T_H \gamma_5 \gamma^\mu {\cal S}({1\over 2}P+q))\Gamma_H(q; P) {\cal S}({1\over 2}P-q)) \big]
\label{fh}
\eeq
The essential quantity is the hadronic matrix element:
\beq
 i \rho^H_\zeta \equiv {- {\langle q \bar q \rangle}^H_\zeta\over f_H} = Z_4 \int^\Lambda \frac{d^4q}{(2\pi)^4} ~
{1\over 2} \big[ T_H\gamma_5  {\cal S}({1\over 2}P+q))\Gamma_H(q; P) {\cal S}({1\over 2}P-q)) \big]
 \label{rhoh}
 \eeq
which takes the place of the usual vacuum expectation value.
Here $T_H$ is a flavor projection operator, $Z_2(\Lambda)$ and $Z_4(\Lambda)$ are renormalization constants, $S(p)$ is the dressed quark propagator, and  $\Gamma_H(q; P)= F.T. \langle H| \psi(x_a) \bar \psi(x_b)|0 \rangle $  is the Bethe-Salpeter bound-state vertex amplitude.
The notation $\langle q \bar q \rangle^H_\zeta$ in the
Bethe-Salpeter analysis thus refers to  a hadronic matrix element, not a vacuum expectation value.
The Bethe-Salpeter analysis of Roberts et al. reproduces the essential features of spontaneous chiral symmetry breaking,  including $m^2_\pi \propto {(m_q +m_{\bar q})/ f_\pi }$ as well as  a finite value for $f_\pi$ at $m_q \to 0.$

One can recast the Bethe Salpeter
formalism into the LF Fock state picture by
time-ordering the coupled BS equation in $\tau = t + z/c$ or by integrating over $dk^-$ where $k^- = k^0 -k^3$ and using
the Wick analysis.  This procedure generates a set of equations which couple
the infinite set of Fock states at fixed $\tau.$ Thus the Casher-Susskind and Bethe-Salpeter
descriptions of spontaneous chiral symmetry breaking and in-hadron condensates and are complementary.

In this paper we show from several physical perspectives that, because of color
confinement, quark and gluon QCD condensates can be regarded as being
associated with the dynamics of hadron wavefunctions, rather than the vacuum
itself.  Thus we analyze the condensates $\langle \bar q q \rangle$ and
$\langle G_{\mu\nu}G^{\mu\nu}\rangle$ and address the question of where they
have spatial (and temporal) support. We argue, in agreement with the original
work of Casher and Susskind~\cite{cs}, that these condensates have spatial
support restricted to the interiors of hadrons, as a consequence of the fact
that they are due to quark and gluon interactions, and these particles are
confined within hadrons.  Higher-order in-hadron condensates such as $\langle
(\bar q q)^2\rangle$, $\langle (\bar q q) G_{\mu\nu}G^{\mu\nu}\rangle$,
etc. are also present, and our discussion implicitly also applies to
these. (The fact that QCD experimentally conserves $P$ and $T$ shows that $P$-
and $T$-noninvariant condensates such as $\langle G_{\mu\nu}\tilde
G^{\mu\nu}\rangle$, where $\tilde G_{\mu\nu} =
(1/2)\epsilon_{\mu\nu\alpha\beta}G^{\alpha\beta}$, are negligible; explaining
this is part of the strong CP problem.) Our analysis includes consideration of
condensed matter analogs, the AdS/CFT correspondence, and the
Bethe-Salpeter/Dyson-Schwinger approach for bound states.  Our analysis
highlights the difference between chiral models where mesons are treated as
elementary fields, and QCD in which all hadrons are composite systems.  We note
that an important consequence of the ``in-hadron'' nature of QCD condensates is
that QCD gives {\it zero} contribution to the cosmological constant, since all
of the gravitational effects of the in-hadron condensate are already included
in the normal contribution from hadron masses.

We emphasize the subtlety in characterizing the formal quantity $\langle 0 |
{\cal O} | 0 \rangle$ in the usual instant form, where ${\cal O}$ is a product
of quantum field operators, by recalling that one can render this automatically
zero by normal-ordering ${\cal O}$ since one has to divide $S$-matrix elements
by vacuum expectation values.  It should be noted that perturbative
contributions to the vacuum in the instant form are not frame independent as
can be seen by computing any bubble diagram -- e.g. in $\phi^3$ theory.  As
shown by Weinberg~\cite{Weinberg:1966jm}, these contributions are suppressed
by powers of $1\over P$ for an observer moving at high momentum $P$.  However,
such contributions are removed by normal-ordering, thus restoring Lorentz
invariance of the instant form vacuum.  Such subtleties are especially delicate
in a confining theory, since the vacuum state in such a theory is not defined
relative to the fields in the Lagrangian, quarks and gluons, but instead
relative to the actual physical, color-singlet, states.

In the front form, the
analysis is simpler, since the physical vacuum is automatically trivial, up to
zero modes.  There are no perturbative bubble diagrams in the LF formalism, so the front-form vacuum is Lorentz invariant from the start.
The light-front method provides a completely consistent formalism
for quantum field theory.  For example, it is straightforward to calculate the
coupling of gravitons to physical particles using the light-front formalism;
in particular, one can prove that the anomalous gravitational magnetic moment
vanishes, Fock state by Fock state~\cite{Brodsky:2000ii}, in agreement with the
equivalence principle~\cite{Teryaev:1999su}. Furthermore, the light-front
method reproduces quantum corrections to the gravitational form factors
computed in perturbation theory~\cite{Berends:1975ah}.

\section{A Condensed Matter Analogy}

 A formulation of quantum field theory using a Euclidean path-integral
(vacuum-to-vacuum amplitude), $Z$, provides a precise meaning for $\langle
{\cal O}\rangle$ as
\beq
\langle {\cal O}\rangle = \lim_{J \to 0} \, \frac{\delta \ln Z}{\delta J} \ ,
\label{oave}
\eeq
where $J$ is a source for ${\cal O}$.  The path integral for QCD, integrated
over quark fields and gauge links using the gauge-invariant lattice
discretization exhibits a formal analogy with the partition function for a
statistical mechanical system.  In this correspondence, a condensate such as
$\langle \bar q q \rangle$ or $\langle G_{\mu\nu}G^{\mu\nu}\rangle$ is
analogous to an ensemble average in statistical mechanics.  To develop a
physical picture of the QCD condensates, we pursue this analogy.  In a
superconductor, the electron-phonon interaction produces a pairing of two
electrons with opposite spins and 3-momenta at the Fermi surface, and, for $T <
T_c$, an associated nonzero Cooper pair condensate $\langle e e
\rangle_T$~\cite{fw}, (here $\langle ... \rangle_T$ means thermal average).
Since this condensate has a phase, the phenomenological Ginzburg-Landau free
energy function
\beq
F = |\nabla \Phi|^2 + c_2 (\Phi^* \Phi) + c_4 (\Phi^* \Phi)^2
\label{flg}
\eeq
uses a complex scalar field $\Phi$ to represent it.  The formal treatment of a
phase transition such as that in a superconductor begins with a partition
function calculated for a finite $d$-dimensional lattice, and then takes the
thermodynamic (infinite-volume) limit.  The non-analytic behavior associated
with the superconducting phase transition only occurs in this infinite-volume
limit; for $T < T_c$, the (infinite-volume) system develops a nonzero value of
the order parameter, namely $\langle \Phi \rangle_T$, in the phenomenological
Ginzburg-Landau model, or $\langle e e \rangle_T$, in the microscopic
Bardeen-Cooper-Schrieffer theory. In the formal statistical mechanics context,
the minimization of the $|{\mathbf \nabla} \Phi|^2$ term implies that the order
parameter is a constant throughout the infinite spatial volume.

However, the infinite-volume limit is an idealization; in reality,
superconductivity is experimentally observed to occur in finite samples of
material, such as Sn, Nb, etc., and the condensate clearly has spatial support
only in the volume of these samples.  This is evident from either of two basic
properties of a superconducting substance, namely, (i) zero-resistance flow of
electric current, and (ii) the Meissner effect, that
\beq
|{\bf B}(z)| \sim |{\bf B}(0)|e^{-z/\lambda_L}
\label{bz}
\eeq
for a magnetic field ${\bf B}(z)$ a distance $z$ inside the superconducting
sample, where the London penetration depth $\lambda_L$ is given by
$\lambda_L^2=m_e c^2/(4\pi n e^2)$, where $n=$ electron concentration; both of
these properties clearly hold only within the sample.  The same statement
applies to other phase transitions such as liquid-gas or ferromagnetic; again,
in the formal statistical mechanics framework, the phase transition and
associated symmetry breaking by a nonzero order parameter at low $T$ occur only
in the thermodynamic limit, but experimentally, one observes the phase
transition to occur effectively in a finite volume of matter, and the order
parameter (e.g., magnetization $M$) has support only in this finite volume,
rather than the infinite volume considered in the formal treatment.  Similarly,
the Goldstone modes that result from the spontaneous breaking of a continuous
symmetry (e.g., spin waves in a Heisenberg ferromagnet) are experimentally
observed in finite-volume samples.  There is, of course, no conflict between
the experimental measurements and the abstract theorems; the key point is that
these samples are large enough for the infinite-volume limit to be a useful
idealization.  Finite-size scaling methods that make this connection precise
are standard tools in studies of phase transitions and critical
phenomena~\cite{finitesize}.

There is another important distinction between condensed matter physics and
relativistic quantum field theories.  The proton eigenstate in QCD is a
summation over Fock states
\beq
\vert P \rangle = \sum_{n=3}^\infty \Psi_{n/P}(x_i, k_{\perp i}, \lambda_i)
\vert n \rangle
\label{LF}
\eeq
where $x_i$ denotes the fraction of the total proton momentum carried by the
parton $i$, $k_{\perp,i}$ denotes the transverse momentum, $\lambda_i$ denotes
the helicity, and the summation extends over states with an unlimited number of
gluons and sea quarks and antiquarks.  In fact, the Regge behavior, $F_2(x,Q^2)
\sim \sum_R \beta_R x^{1-\alpha_R}$, of hadronic structure functions
at small $x$ requires that the hadronic wavefunction has Fock states $\vert n
\rangle$ with an infinite number of quark and gluon partons.  (Here, in
standard notation, $x = -q^2/(2M_N\nu)$, where $\nu$ denotes the energy
transfer; $\beta_R$ denotes the amplitude with which a Regge trajectory
contributes to the scattering, and $\alpha_R$ denotes the intercept of this
trajectory.) This relation applies in the Regge region, $s >> \Lambda_{QCD}^2$
with $t=-q^2$ fixed, i.e., small $x$. For example,
Mueller~\cite{Mueller:1993rr} has shown that the BFKL
(Balitsky-Fadin-Kuraev-Lipatov) behavior of the structure functions at $x \to
0$ is a result of the infinite range of gluonic Fock states. The relation
between Fock states of different $n$ is given by an infinite tower of ladder
operators~\cite{Antonuccio:1997tw}.  In the analysis by Casher and
Susskind~\cite{cs}, spontaneous chiral symmetry breaking occurs within the
confines of the hadron wavefunction due to a phase change supported by the
infinite number of quark and quark pairs in the light-front wavefunction. Thus,
as noted above, unlike the usual discussion in condensed matter physics,
infinite volume is not required for a phase transition in relativistic quantum
field theories.

\section{A Picture of QCD Condensates}

The condensed-matter physics discussion above helps to motivate our analysis
for QCD.  The spatial support for QCD condensates should be where the particles
are whose interactions give rise to them, just as the spatial support of a
magnetization $M$, say, is inside, not outside, of a piece of iron.  The
physical origin of the $\langle \bar q q \rangle$ condensate in QCD can be
argued to be due to the reversal of helicity (chirality) of a massless quark as
it moves outward and reverses its three-momentum at the boundary of a hadron
due to confinement \cite{casher}.  This argument suggests that the condensate
has support only within the spatial extent where the quark is confined; i.e.,
the physical size of a hadron.  Another way to motivate this is to note that in
the light-front Fock state picture of hadron
wavefunctions~\cite{cs,bpprev,bs94}, a valence quark can flip its chirality
when it interacts or interchanges with the sea quarks of multiquark Fock
states, thus providing a dynamical origin for the quark running mass.  In this
description, the $\langle \bar q q \rangle$ and $\langle
G_{\mu\nu}G^{\mu\nu}\rangle$ condensates are effective quantities which
originate from $q \bar q$ and gluon contributions to the higher Fock state
light-front wavefunctions of the hadron and hence are localized within the
hadron.  There is a natural relation with the nucleon sigma term, $\sigma_{\pi
N} = (1/2)(m_u+m_d)\langle N | \bar q q | N \rangle$ (where here the nucleon
states are normalized as $\langle N(p')|N(p)\rangle = (2\pi)^3\delta^3({\mathbf
p} -{\mathbf p}')$). As discussed in the introduction, the vacuum condensate appearing in the 
Gell Mann-Oakes-Renner relation~\cite{gmor}
\beq
m_\pi^2 = -\frac{(m_u+m_d)}{f_\pi^2} \, \langle \bar q q \rangle
\label{gmor}
\eeq
is replaced by the in-hadron condensate, as defined in Eq. (\ref{rhoh}).

\section{Chiral Symmetry Breaking in the AdS/CFT Model}

The Anti-De Sitter/conformal field theory (AdS/CFT) correspondence between
string theory in AdS space and CFT's in physical spacetime has been used to
obtain an analytic, semi-classical model for strongly-coupled QCD which has
scale invariance and dimensional counting at short distances and color
confinement at large distances~\cite{lightfront}.  Color confinement can be
imposed by introducing hard-wall boundary conditions at $z={1/ \Lambda_{QCD}}$
($z=$ AdS fifth dimension) or by modification of the AdS metric.  This AdS/QCD
model gives a good representation of the mass spectrum of light-quark mesons
and baryons as well as the hadronic wavefunctions~\cite{lightfront}.  One can
also study the propagation of a scalar field $X(z)$ as a model for the
dynamical running quark mass~\cite{lightfront}. The AdS solution has the
form~\cite{xz}
\beq
X(z) = a_1 z+ a_2 z^3 \ ,
\label{xz}
\eeq
where $a_1$ is proportional to the current-quark mass. The coefficient $a_2$
scales as $\Lambda^3_{QCD}$ and is the analog of $\langle \bar q q \rangle$;
however, since the quark is a color nonsinglet, the propagation of $X(z),$ and
thus the domain of the quark condensate, is limited to the region of color
confinement.  The AdS/QCD picture of effective confined condensates is in
agreement with results from chiral bag models~\cite{chibag}, which modify the
original MIT bag by coupling a pion field to the surface of the bag in a
chirally invariant manner. Since the effect of $a_2$ depends on $z$, the AdS
picture is inconsistent with the usual picture of a constant condensate.

\section{Empirical Determinations of the Gluon Condensate}

The renormalization-invariant quantity $\langle (\alpha_s/\pi)
G_{\mu\nu}G^{\mu\nu} \rangle$, where
\beq
G_{\mu\nu}G^{\mu\nu} = 2\sum_a (|{\bf B}^a|^2 - |{\bf E}^a|^2)) \ ,
\label{gbe}
\eeq
can be determined empirically by analyzing vacuum-to-vacuum current correlators
constrained by data for $e^+e^- \to$ charmonium and hadronic $\tau$
decays~\cite{svz}-\cite{ggval}. (Here we use units where $\hbar=c=1$, and our
flat-space metric is $\eta_{\mu\nu}={\rm diag}(1,-1,-1,-1)$).  Some recent
values (in GeV$^4$) include $0.006 \pm 0.012$~\cite{ggval}(a), $0.009\pm
0.007$~\cite{ggval}(b), and $-0.015 \pm 0.008$~\cite{ggval}(c).  These values
show significant scatter and even differences in sign. These are consistent
with the picture in which the vacuum gluon condensate vanishes; it is confined
within hadrons, rather than extending throughout all of space, as would be true
of a vacuum condensate.

\section{Some Other Features of QCD Condensates}

In the picture discussed here, QCD condensates would be considered as
contributing to the masses of the hadrons where they are located.  This is
clear, since, e.g., a proton subjected to a constant electric field will
accelerate and, since the condensates move with it, they comprise part of its
mass. Similarly, when a hadron decays to a non-hadronic final state, such as
$\pi^0 \to \gamma\gamma$, the condensates in this hadron contribute their
energy to the final-state photons. Thus, over long times, the dominant regions
of support for these condensates would be within nucleons, since the proton is
effectively stable (with lifetime $\tau_p >> \tau_{univ} \simeq 1.4 \times
10^{10}$ yr.), and the neutron can be stable when bound in a nucleus.  In a
process like $e^+e^- \to$ hadrons, the formation of the condensates occurs on
the same time scale as hadronization.  In accord with the Heisenberg
uncertainty principle, these QCD condensates also affect virtual processes
occurring over times $t \lsim 1/\Lambda_{QCD}$.

Moreover, in our picture, condensates $\langle \bar q q \rangle$ in different
hadrons may be chirally rotated with respect to each other, somewhat analogous
to disoriented chiral condensates in heavy-ion collisions~\cite{dcc}.  This
picture of condensates can, in principle, be verified by careful lattice gauge
theory measurements.  Note that the lattice measurements that have inferred
nonzero values of $\langle \bar q q \rangle$ were performed in finite volumes,
although these were usually considered as approximations to the infinite-volume
limit (for an early review of $\langle \bar q q \rangle$ lattice measurements,
see \cite{kogut}; recent reviews are in \cite{lat08}).

\section{Application to Other Asymptotically Free Gauge Theories}

Having discussed QCD, we next consider, as an exercise, how this approach to
condensates would apply to several hypothetical asymptotically free gauge
theories.  We begin with a vectorial gauge theory with the gauge group
SU($N_c$), allowing $N_c$ to be generalized to values $N_c \ge 3$.  First,
consider a theory of this type with no fermions, so that only $\langle
G_{\mu\nu}G^{\mu\nu} \rangle$ need be considered.  This condensate would then
have support within the interior of the glueballs. Second, consider a theory
with $N_f=1$ massless or light fermion transforming according to some
nonsinglet representation $R$ of SU($N_c$).  The $\langle \bar q q \rangle$ and
$\langle G_{\mu\nu}G^{\mu\nu}\rangle$ condensates in this theory would have
support in the interior of the mesons, baryons, and glueballs (or mass
eigenstates formed from glueballs and mesons).  Here, the condensate $\langle
\bar q q \rangle$ does not break any non-anomalous global chiral symmetry, so
there would not be any Nambu-Goldstone boson (NGB).  In both of these theories,
the sizes of the mesons, baryons, and glueballs are $\simeq 1/\Lambda$, where
$\Lambda$ is the confinement scale.  Another application would be to a strongly
coupled vectorial gauge interaction that could possibly play a role in
dynamical electroweak symmetry breaking \cite{tc}.

We next consider asymptotically free chiral gauge theories (which are free of
gauge and global anomalies) with massless fermions transforming as
representations $\{R_i\}$ of the gauge group. The properties of strongly
coupled theories of this type are not as well understood as those of vectorial
gauge theories \cite{thooft}-\cite{tum}.  One possibility is that, as the
energy scale decreases from large values and the associated running coupling
$g$ increases, it eventually becomes large enough to produce a (bilinear)
fermion condensate, which thus breaks the initial gauge symmetry
\cite{tum}. This is expected to form in the most attractive channel (MAC) $R_1
\times R_2 \to R_{cond.}$, which maximizes the quantity $\Delta C_2 =
C_2(R_1)+C_2(R_2) - C_2(R_{cond.})$, where $C_2(R)$ is the quadratic Casimir
invariant.  Depending on the theory, several stages of self-breaking may occur
\cite{tum,ckm}.  Let us consider an explicit model of this type, with gauge
group SU(5) and massless left-handed fermion content consisting of an
antisymmetric rank-2 tensor representation, $\psi^{ij}_L$, and a conjugate
fundamental representation, $\chi_{i,L}$.  This theory is asymptotically free
and has a formal ${\rm U}(1)_\psi \times {\rm U}(1)_\chi$ global chiral
symmetry; both U(1)'s are broken by SU(5) instantons, but the linear
combination U(1)$^\prime$ generated by $Q = Q_\psi -3Q_\chi$ is preserved.  The
MAC for condensation is
\beq
10 \times 10 \to \bar 5
\label{10105bar}
\eeq
with $\Delta C_2=24/5$, and
the associated condensate is
\beq
\langle \epsilon_{ijk\ell n}\psi^{jk \ T}_L C \psi^{\ell n}_L \rangle \ ,
\label{cond10105bar}
\eeq
which breaks SU(5) to SU(4).  Thus, as the energy scale decreases and the
running $\alpha=g^2/(4\pi)$ grows, at a scale $\Lambda$ at which $\alpha \Delta
C_2 \sim O(1)$, this condensate is expected to form. Without loss of
generality, we take $i=1$, and note
\beq
\langle \epsilon_{1jk\ell n}\psi^{jk \ T}_L C \psi^{\ell n}_L \rangle
\propto
\langle \psi^{23 \ T}_L C \psi^{45}_L - \psi^{24 \ T}_L C \psi^{35}_L
 + \psi^{25 \ T}_L C \psi^{34}_L \rangle
\label{1010condensate}
\eeq
The nine gauge bosons in the coset SU(5)/SU(4) gain masses of order
$\Lambda$. The six components of $\psi^{ij}_L$ involved in the condensate
(\ref{1010condensate}) also gain dynamical masses of order $\Lambda$. These
components bind to form an SU(4)-singlet meson whose wavefunction is given by
the operator in (\ref{1010condensate}).  This binding involves the exchange of
the various (perturbatively massless) gauge bosons of SU(4).  The condensate
(\ref{1010condensate}) breaks the global U(1)$^\prime$, but the would-be
resultant NGB is absorbed by the gauge boson corresponding to the diagonal
generator in SU(5)/SU(4).  According to the picture discussed here, the
condensate (\ref{1010condensate}) would have spatial support in the meson with
the same wavefunction.  Aside from the SU(4)-singlet $\chi_{1,L}$, the
remaining massless fermion content of the SU(4) theory is vectorial, consisting
of a 4, $\psi^{1j}_L$, and a $\bar 4$, $\chi_{j,L}$, $j=2...4$.  The formal
global flavor symmetry of this effective SU(4) theory at energy scales below
$\Lambda$ is
\beq
{\rm U}(1)_L \times {\rm U}(1)_R = {\rm U}(1)_V \times {\rm U}(1)_A \ ,
\label{chirsym}
\eeq
and the U(1)$_A$ is broken by SU(4) instantons.  This low-energy
effective field theory is asymptotically free, so that at lower energy scales,
the coupling $\alpha$ that it inherits from the SU(5) theory continues to
increase, and the theory confines and produces the condensate $\langle \psi^{1j
\ T}_L C \chi_{j,L} \rangle$, which preserves the gauged SU(4) and global
U(1)$_V$.  We infer that $\langle \psi^{1j \ T}_L C \chi_{j,L} \rangle$ and the
SU(4) gluon condensate $\langle G_{\mu\nu}G^{\mu\nu}\rangle$ have spatial
support in the SU(4)-singlet baryon, meson, and glueball states of this theory.

Although the present picture associates condensates in a confining gauge theory
$G$ with $G$-singlet hadrons, these condensates can affect properties of
$G$-singlet particles if they both couple to a common set of fields.  For
example, the $\langle \bar F F \rangle$ condensate and the corresponding
dynamical mass $\Sigma_F$ of technifermions in a technicolor (TC) theory give
rise to the masses of the (TC-singlet) quarks and leptons via diagrams
involving exchanges of virtual extended technicolor gauge bosons. Our analysis
could also be extended to supersymmetric gauge theories, but we shall not
pursue this here.

\section{The Case of an Infrared-Free Gauge Theory}

Our discussion is only intended to apply to asymptotically free gauge theories.
However, we offer some remarks on the situation for a particular infrared-free
theory here, namely a U(1) gauge theory with gauge coupling $e$ and some set of
fermions $\psi_i$ with charges $q_i$. Here there are several important
differences with respect to an asymptotically free non-Abelian gauge theory.
First, while the chiral limit of QCD, i.e., quarks with zero current-quark
masses, is well-defined because of quark confinement, a U(1) theory with
massless charged particles is unstable, owing to the well-known fact that these
would give rise to a divergent Bethe-Heitler pair production cross section.  It
is therefore necessary to break the chiral symmetry explicitly with bare
fermion mass terms $m_i$.  If the running coupling $\alpha_1=e^2/(4\pi)$ at a
given energy scale $\mu$ were sufficiently large, $\alpha_1(\mu) \gsim O(1)$,
an approximate solution to the Schwinger-Dyson equation for the propagator of a
fermion $\psi_i$ with $m_i << \mu$ would suggest that this fermion gains a
nonzero dynamical mass $\Sigma_i$ \cite{sd} and hence, presumably, there would
be an associated condensate $\langle \bar \psi_i \psi_i\rangle$ (no sum on
$i$).  However, in analyzing S$\chi$SB, it is important to minimize the effects
of explicit chiral symmetry breaking due to the bare masses $m_i$. The
infrared-free nature of this theory means that for any given value of
$\alpha_1$ at a scale $\mu$, as one decreases $m_i/\mu$ to reduce explicit
breaking of chiral symmetry, $\alpha_1(m_i)$ also decreases, approaching zero
as $m_i/\mu \to 0$.  Since $\alpha_1(m_i)$ should be the relevant coupling to
use in the Schwinger-Dyson equation, it may in fact be impossible to realize a
situation in this theory in which one has small explicit breaking of chiral
symmetry and a large enough value of $\alpha_1(m_i)$ to induce spontaneous
chiral symmetry breaking.  A full analysis would require knowledge of the bound
state spectrum of the hypothetical strongly coupled U(1) theory, but this
spectrum is not reliably known.

\section{Finite Temperature QCD}

So far, we have discussed QCD and other theories at zero temperature (and
chemical potential or equivalently here, baryon density).  For QCD in thermal
equilibrium at a finite temperature $T$, as $T$ increases above the
deconfinement temperature $T_{dec}$, both the hadrons and the associated
condensates eventually disappear, although experiments at CERN and BNL-RHIC
show that the situation for $T \gsim T_{dec}$ is more complicated than a weakly
coupled quark-gluon plasma.  The picture of the QCD condensates here is
especially close to experiment, since, although finite-temperature QCD makes
use of the formal thermodynamic, infinite-volume limit, actual heavy ion
experiments and resultant transitions from confined to deconfined quarks and
gluons take place in the finite volume and time interval provided by colliding
heavy ions.  Indeed, one of the models that has been used to analyze such
experiments involves the notion of a color-glass condensate \cite{cgc}.

\section{QCD and the Cosmological Constant}

One of the most challenging problems in physics is that of the cosmological
constant $\Lambda$ (recent reviews include \cite{peeblesratra}-\cite{detf}).
This enters in the Einstein gravitational field equations as \cite{bks} 
\beq
R_{\mu\nu}-\frac{1}{2}g_{\mu\nu}R-\Lambda g_{\mu\nu} = (8\pi G_N)T_{\mu\nu},
\label{einsteineq}
\eeq
where $R_{\mu\nu}$, $R$, $g_{\mu\nu}$, $T_{\mu\nu}$, and $G_N$ are the Ricci
curvature tensor, the scalar curvature, the metric tensor, the stress-energy
tensor, and Newton's constant. One defines
\beq
\rho_\Lambda = \frac{\Lambda}{8 \pi G_N}
\label{rholambda}
\eeq
and
\beq
\Omega_\Lambda = \frac{\Lambda}{3H_0^2} = \frac{\rho_\Lambda}{\rho_c}~,
\label{omegalambda}
\eeq
where
\beq
\rho_c = \frac{3H_0^2}{8\pi G_N} \ ,
\label{rhoc}
\eeq
and $H_0=(\dot a/a)_0$ is the Hubble constant in the present era, with $a(t)$
being the Friedmann-Robertson-Walker scale parameter \cite{bks}.  The field
equations imply $(\dot a/a)^2 = H^2 = (8\pi G_N/3)\rho + \Lambda/3 - k/a^2$ and
$\ddot a/a = -4\pi G_N (\rho+3p)+\Lambda/3$, where $\rho=$ total mass/energy
density, $p=$ pressure, and $k$ is the curvature parameter; equivalently, $1 =
\Omega_m + \Omega_\gamma + \Omega_\Lambda + \Omega_k$, where $\Omega_m = 8\pi
G_N \rho_m/H_0^2$, $\Omega_\gamma = 8\pi G_N \rho_\gamma/H_0^2$, and $\Omega_k
= -k/(H_0^2 a^2)$.  Long before the current period of precision cosmology, it
was known that $\Omega_\Lambda$ could not be larger than O(1). In the context
of quantum field theory, this was very difficult to understand, because
estimates of the contributions to $\rho_\Lambda$ from (i) vacuum condensates of
quark and gluon fields in QCD and the vacuum expectation value of the Higgs
field hypothesized in the Standard Model (SM) to be responsible for electroweak
symmetry breaking, and from (ii) zero-point energies of quantum fields appear
to be too large by many orders of magnitude.  Observations of supernovae showed
the accelerated expansion of the universe and are consistent with the
hypothesis that this is due to a cosmological constant, $\Omega_\Lambda \simeq
0.76$ \cite{hzst,scp}.  The supernovae data \cite{hzst,scp}, together with
measurements of the cosmic microwave background radiation, galaxy clusters, and
other inputs, e.g., primordial element abundances, have led to a consistent
determination of the cosmological parameters \cite{wmap}-\cite{pdg}.  These
include $H_0 = 73 \pm 3$ km/s/Mpc, $\rho_c = 0.56 \times 10^{-5}$ GeV/cm$^3$ $=
(2.6 \times 10^{-3} \ {\rm eV} \,)^4$, total $\Omega_m \simeq 0.24$ with baryon
term $\Omega_b \simeq 0.042$, so that the dark matter term is $\Omega_{dm}
\simeq 0.20$.  In the equation of state $p=w\rho$ for the ``dark energy'', $w$
is consistent with being equal to $-1$, the value if the accelerated expansion
is due to a cosmological constant.  (Other suggestions for the source of the
accelerated expansion include modifications of general relativity and
time-dependent $w(t)$, as reviewed in \cite{peeblesratra}-\cite{detf}.)

Here, using our observations concerning QCD condensates, we propose a solution
to problem (i) of the contributions by these condensates to $\rho_\Lambda$,
which, in the conventional approach, are much too large. The QCD
condensates form at times of order $10^{-5}$ sec. in the early universe, as the
temperature $T$ decreases below the confinement-deconfinement temperature
$T_{dec} \simeq 200$ MeV.  As noted above, for $T << T_{dec}$, in the
conventional quantum field theory view, these condensates are considered to be
constants throughout space.  If one accepts this conventional view, then these
condensates would contribute $(\delta \rho_\Lambda)_{QCD} \sim
\Lambda_{QCD}^4$, so that $(\delta \Omega_\Lambda)_{QCD} \simeq 10^{45}$. On
the contrary, if one accepts the argument that these condensates (and also
higher-order ones such as $\langle (\bar q q)^2\rangle$ and $\langle (\bar q q)
G_{\mu\nu}G^{\mu\nu}\rangle$) have spatial support within hadrons, not
extending throughout all of space, then one makes considerable progress in
solving the above problem, since the effect of these condensates on gravity is
already included in the baryon term $\Omega_b$ in $\Omega_m$ and, as such, they
do not contribute to $\Omega_\Lambda$.

Another excessive type-(i) contribution to $\rho_\Lambda$ is conventionally
viewed as arising from the vacuum expectation value of the Standard-Model Higgs
field, $v_{EW} = 2^{-1/4}G_F^{-1/2} = 246$ GeV, giving $(\delta
\rho_\Lambda)_{EW} \sim v_{EW}^4$ and hence $(\delta \Omega_\Lambda)_{EW} \sim
10^{56}$.  Similar numbers are obtained from Higgs vacuum expectation values in
supersymmetric extensions of the Standard Model (recalling that the
supersymmetry breaking scale is expected to be the TeV scale).  However, it is
possible that electroweak symmetry breaking is dynamical; for example, it may
result from the formation of a bilinear condensate of fermions $F$ (called
technifermions) subject to an asymptotically free, vectorial, confining gauge
interaction, commonly called technicolor (TC), that gets strong on the TeV
scale \cite{tc}. In such theories there is no fundamental Higgs field.
Technicolor theories are challenged by, but may be able to survive, constraints
from precision electroweak data.  By our arguments above, in a
technicolor theory, the technifermion and technigluon condensates would have
spatial support in the technihadrons and techniglueballs and would contribute
to the masses of these states.  We stress that, just as was true for the QCD
condensates, these technifermion and technigluon condensates would not
contribute to $\rho_\Lambda$.  Hence, if a technicolor-type mechanism should
turn out to be responsible for electroweak symmetry breaking, then there would
not be any problem with a supposedly excessive contribution to $\rho_\Lambda$
for a Higgs vacuum expectation value.  Indeed, stable technihadrons in certain
technicolor theories may be viable dark-matter candidates. 

We next comment briefly on type-(ii) contributions. The formal expression for
the energy density $E/V$ due to zero-point energies of a quantum field
corresponding to a particle of mass $m$ is
\beq
E/V = \int \frac{d^3k}{(2\pi)^3} \frac{\omega(k)}{2} \ ,
\label{zeropoint}
\eeq
where the energy is $\omega(k) = \sqrt{{\bf k}^2 + m^2}$.  However, first, this
expression is unsatisfactory, because it is (quadratically) divergent.  In
modern particle physics one would tend to regard this divergence as indicating
that one is using an low-energy effective field theory, and one would impose an
ultraviolet cutoff $M_{UV}$ on the momentum integration, reflecting the upper
range of validity of this low-energy theory.  Since neither the left- nor
right-hand side of eq. (\ref{zeropoint}) is Lorentz-invariant, this cutoff
procedure is more dubious than the analogous procedure for Feynman integrals of
the form $\int d^4k \, I(k,p)$ in quantum field theory, where
$I(k,p_1,...,p_n)$ is a Lorentz-invariant integrand function depending on some
set of 4-momenta $p_1,...,p_n$.  If, nevertheless, one proceeds to use such a
cutoff, then, since a mass scale characterizing quantum gravity (QG) is
$M_{Pl}=G_N^{-1/2}=1.2 \times 10^{19}$ GeV, one would infer that $(\delta
\rho_\Lambda)_{QG} \sim M_{Pl}^4/(16\pi^2)$, and hence $(\delta
\Omega_\Lambda)_{QG} \sim 10^{120}$.  With the various mass scales
characterizing the electroweak symmetry breaking and particle masses in the
Standard Model, one similarly would obtain $(\delta \Omega_\Lambda)_{SM} \sim
10^{56}$.  Given the fact that eq. (\ref{zeropoint}) is not Lorentz-invariant,
one may well question the logic of considering it as a contribution to the
Lorentz-invariant quantity $\rho_\Lambda$. (This criticism of the conventional
lore has also been made in \cite{peeblesratra} and \cite{jaffe05}.) Indeed, one
could plausibly argue that, as an energy density, it should instead be part of
$T_{00}$ in the energy-momentum tensor $T_{\mu\nu}$.  Phrased in a different
way, if one argues that it should be associated with the $\Lambda g_{\mu \nu}$
term, then there must be a negative corresponding zero-point pressure
satisfying $p=-\rho$, but the source for such a negative pressure is not
evident in eq. (\ref{zeropoint}).  The light-front approach to the construction
of a quantum field theory, in particular, the Standard
Model, provides another perspective to this issue \cite{lf}-\cite{scottish}.

\section{Concluding Remarks}

We have argued from several physical perspectives that, because of color
confinement, quark and gluon QCD condensates are localized within the interiors
of hadrons. Our analysis is in agreement with the Casher and Susskind model and
the explicit demonstration of ``in-hadron condensates" by Roberts et al., using
the Bethe-Salpeter/Dyson-Schwinger formalism for QCD bound states.  We also
discussed this physics using condensed matter analogues, the AdS/CFT
correspondence, and the Bethe-Salpeter/Dyson-Schwinger approach for
bound states.

In-hadron condensates provide a  solution to what has hitherto commonly been
regarded as an excessively large contribution to the cosmological constant by
QCD condensates. We have argued that these condensates do not, in fact,
contribute to $\Omega_\Lambda$; instead, they have spatial support within
hadrons and, as such, should really be considered as contributing to the masses
of these hadrons and hence to $\Omega_b$. We have also suggested a possible
solution to what would be an excessive contribution to $\Omega_\Lambda$ from a
hypothetical Higgs vacuum expectation value; the solution would be applicable
if electroweak symmetry breaking occurs via a technicolor-type mechanism.

\bigskip
\bigskip

We thank R. Alkofer, A. Casher, C. Fischer, M. E. Fisher, F. Llanes-Estrada,
C. Roberts, L. Susskind, P. Tandy, and G. F. de T\'eramond for helpful
conversations. This research was partially supported by grants
DE-AC02-76SF00515 (SJB) and NSF-PHY-06-53342 (RS).

\end{document}